\documentstyle[psfig,conf_iap,10pt]{article}
\def\plotfiddle#1#2#3#4#5#6#7{\centering \leavevmode
    \vbox to#2{\rule{0pt}{#2}}
    \includegraphics{#1}}
\begin{document}
\heading{%
Effects of Galaxy Selection Upon Ly$\alpha$ Absorber \\ Identification
} 
\par\medskip\noindent
\author{%
Suzanne M. Linder$^1$
}
\address{%
The Pennsylvania State University, Department of Astronomy and Astrophysics,
525 Davey Laboratory, University Park, PA 16802, USA
}

\begin{abstract}
While it is possible to explain Ly$\alpha$ absorber counts at low
redshift using gas which is associated with moderately extended galaxies
\cite{sml97}, absorbers are often observed to be associated with galaxies at
larger impact parameters from quasar lines of sight than are expected from
calculated galaxy absorption cross sections in such absorber-galaxy models.
However, a large fraction of absorbers is expected to arise in lines sight
through galaxies which are low in luminosity and/or surface brightness, so
that they are unlikely to be detected in surveys for galaxies close to quasar
lines of sight.  Given that it is impossible to be certain that any particular
absorber has been matched to the correct galaxy, I show that it is possible
to simulate plots of absorption covering factors around luminous galaxies
which resemble observed plots by assuming that absorption often originates
in unidentified galaxies.
\end{abstract}

Supposing that Ly$\alpha$ absorbers at low redshift generally arise from
lines of sight through galaxies, it is easily possible
to explain absorber counts when including absorption from dwarf and low
surface brightness galaxies.  Furthermore, it is possible to put an upper limit
on the characteristic absorbing cross section of galaxies by comparing results
from simulations with observed galaxy luminosity functions.
For example, a galaxy with $M_B^*=-18.9$ is likely to have an absorbing
radius of around 200-300 kpc assuming that absorption arises in
extended disks \cite{sml97}.
Surveys for luminous galaxies around quasar lines of sight have found
possible absorption at much larger impact parameters from such galaxies
\cite{lebrun96}, \cite{morris93}, and surveys which include fainter
galaxies \cite{bowen96}, \cite{lanzetta95} also find large absorption covering 
factors at large galaxy impact parameters compared to simulation results
\cite{sml97}.  All surveys are limited in galaxy magnitude, and 
few attempts have been made to search for low surface brightness galaxies
around quasar lines of sight \cite{vangorkom96}.  A large fraction of absorbers
may originate in galaxies which are low in luminosity and/or surface 
brightness, so that they would not usually be detected in these surveys.
Thus the absorption lines are likely to be observationally matched with other 
galaxies.

%
\begin{figure}[htb]
\plotfiddle{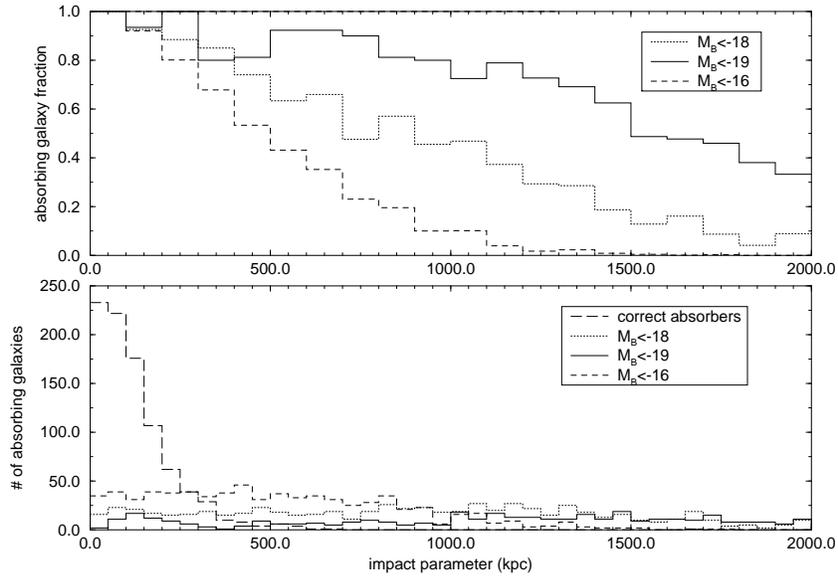}{2.5in}{270}{40}{40}{-164}{250}
\vskip -0.4in
\caption{(a) The fraction of galaxies ($\mu_B(0)<22$ mag arcsec$^{-2}$)
within 400 km/s of an absorption line,
which are identified as causing absorption ($>10^{14.3}$ cm$^{-2}$) is plotted 
versus the galaxy impact 
parameter from the line of sight.  (b) The impact parameters for the correct
absorbing galaxies are plotted along with those identified as 
causing absorption within the given magnitude limits.  No galaxy is identified
within 2 Mpc of the line of sight for 2\% and 10\% of the absorbers for 
$M_B<-18$ and $-19$ respectively.
}
\end{figure}

The simulation \#10 described in \cite{sml97} was repeated with the following
adjustment:  In order to make the number density of galaxies more realistic
according to absorber counts, 31500 galaxies were simulated and placed in
a cube with side length of 25 Mpc.  Rather than matching each absorber with 
the known responsible galaxy from the simulation, the nearest galaxy was 
found for each absorber which satisfied some selection criteria.  'Covering
factor' plots are shown in Figure 1a, which give the fraction of galaxies 
located in each range of impact parameter which are found to cause absorption.
Here many galaxies appear to cause absorption at larger impact parameters 
than those in Figure 7 of \cite{sml97}.  The distribution of absorbing 
galaxy impact parameters is shown in Figure 1b for the galaxies which are 
matched to absorbers according to the selection criteria and for the correct
absorbing galaxies known from the simulations.  Note that a galaxy which 
satisfies the selection criteria is often not found within 2 Mpc of a line
of sight.
Many absorbers are likely to arise from low surface
brightness galaxies which are clustered around more luminous objects.  Thus
simulating accurate covering factor plots will require including
clustering of galaxies in the models.  When the selection procedures used
to identify absorbing galaxies are taken into account
the absorber-galaxy models described here may be able to produce
realistic covering factor plots.

\vfill
\end{document}